\documentclass[aps,pra,floatfix,twocolumn,showpacs]{revtex4}
\usepackage{dcolumn}

\begin{document}
\title{A New Option for a Search for $\alpha$ Variation:
Narrow Transitions with Enhanced Sensitivity}

\author{S. G. Karshenboim}
\affiliation{ D. I. Mendeleev Institute for Metrology
(VNIIM),
St. Petersburg 198005, Russia}
\affiliation
{Max-Planck-Institut f\"ur Quantenoptik, 85748
Garching, Germany
}

\author{A. Yu. Nevsky }
\affiliation{ Institut f\"ur Experimentalphysik,
Heinrich-Heine-Universit\"at D\"usseldorf,  D-40225 D\"usseldorf,
Germany}
\author{E. J. Angstmann$^1$, V. A. Dzuba$^1$}
\author{V. V. Flambaum$^1$$,$$^{2}$}
\affiliation{$^1$School of Physics, University of New South Wales, Sydney 2052,
Australia\\
$^2$ Physics Division, Argonne National Laboratory, Argonne,
Illinois 60439-4843, USA }

\date{\today}


\begin{abstract}
We consider several transitions between narrow lines that have an
enhanced sensitivity to a possible variation of the fine structure
constant, $\alpha$. This enhancement may allow
 a search to be performed with an effective suppression of the systematic
sources of uncertainty that are unavoidable in conventional high-resolution
spectroscopic measurements. In the future this may provide the strongest
laboratory constraints on $\alpha$ variation.
\end{abstract}

\maketitle

\section{Introduction}

Over the last decade there has been an increasing interest in a possible
time-variation of fundamental physical constants (see e.g.
\cite{acfcbook}). This stimulated a number of theoretical
speculations and high-precision experiments aiming to analyze
various models and search for such phenomena. There have been
several different motivations for such studies, ranging from theories which
claim to better describe certain deeply fundamental features of nature,
to the development of a new generation of ultra-stable clocks.
 Fundamental physics suggests that our Universe has already
experienced one or a few phase transitions during its evolution with
dramatic changes to the mass of leptons and current quarks, the
fine structure constant, $\alpha$, etc. In addition to this, certain models of the
unification of the electroweak and strong interactions with gravity,
or even attempts at the development of a quantum theory of gravity
alone, may involve time and/or space variations of the base
fundamental constants. On the experimental front, a recent breakthrough in
frequency metrology, including the achievement of a record accuracy in
microwave fountain standards, and the development of a new
generation of optical clocks, requires strong practical tests. A cross
comparison of several frequency standards serves both purposes:
a search for a possible variation of the fundamental constants, that could
point towards new physics, and a routine check of the
most advanced frequency standards that are important even for our
everyday life, such as through various high-precision global
navigation systems.

Unification schemes, cosmological models, and quantum theories of
gravity indicate that certain variations to the value of certain
fundamental physical constants are possible but, unfortunately,
they can not supply us with quantitative details. Neither the
hierarchy of the expected variation rates of the different
fundamental constants is understood, nor is the form of the dependence clear.
 Is it a space or time variation? Is it a nearly linear
drift or does it oscillate? Without responding to these basic questions
we are not in a position to find out what is the most sensitive way
to search for possible variations.

A comparison of different experiments to verify their results and
check their consistency is also far from an easy solution. In
particular a few kinds of the results have been obtained up to now
and may be improved upon in the near future.
\begin{itemize}
\item Astrophysical observations of absorption spectra of quasars
have delivered questionable results. A positive indication of a
variation to the fine structure constant
\begin{equation}
\frac{\Delta \alpha}{\alpha} =\bigl(-0.54\pm0.12\bigr)
\cdot10^{-5}
\end{equation}
at the 5 sigma level, associated with red shifts in the range $0.2
< z < 3.7$, which corresponds to a time separation of $2.5 - 12.2$
Gyr in the currently popular model with $\Omega_{\Lambda}=0.7$,
$\Omega_{\rm matter}=0.3$, and $H_{0} = 68$ kms$^{-1}$Mpc$^{-1}$,
was obtained \cite{2,3,4}.
Meanwhile the recent results of other groups are consistent with
zero variation at the same level of accuracy, ${\Delta
\alpha}/{\alpha} = (-0.06 \pm 0.06)\cdot10^{-5}$ at
 $ 0.4 < z < 2.3 $ ($4.4 - 11.0$ Gyr)
\cite{astro_zero} and ${\Delta \alpha}/{\alpha} = (-0.04\pm
0.33)\cdot10^{-5}$ at $z\simeq 1.15$ corresponding to $8.5$ Gyr
ago \cite{astro_zero1}. All three evaluations are based on the
so-called many-multiplet method suggested in \cite{dzubaprl}
 (a modification of this method
 was applied in \cite{astro_zero1}) and are related to approximately the same
red-shift. However, works \cite{astro_zero,astro_zero1} use data
from a different telescope observing a different (Southern)
hemisphere.
\item An initial negative result from the Oklo uranium mine
\cite{oklo}, related to a variation of a samarium resonance in the
100-meV range of the thermal neutron absorption spectrum about
$2\times10^9$ years ago, has probably changed to a positive signal
\cite{steve}. An upper boundary at the same level of accuracy was
also achieved in study of slow radioactive decays \cite{decay}.
The results involve numerous assumptions at various stages of
examination, and in particular, when the result for a variation of
a non-fundamental quantity is turned into terms of $\alpha$
variation. Actually, the Oklo and  radioactive decay results are
more sensitive to the variation of the strong interaction, rather
than to the variation of $\alpha$.
\item A negative result from a comparison of rubidium to cesium
hyperfine splitting \cite{paris}
 corresponds to a
constraint on a time variation of the proton $g$ factor, rather
than of $\alpha$ variation (see, e.g., \cite{cjp}).
\item Recent optical measurements of transitions in the gross
structure of neutral calcium and hydrogen, and singly charged ions
of mercury and ytterbium, set a constraint on possible $\alpha$
variation to the level of few parts in $10^{15}$ per year
\cite{newYb,amo}
\begin{equation}
\frac{\partial \ln \alpha}{\partial t}=(-0.3\pm 2.0)\cdot
10^{-15}\,{\rm yr}^{-1}\;.
\end{equation}
\end{itemize}
The experiments and observations mentioned above are related to
different time intervals and there is no reliable method for their
model-independent comparison. Actually the original results
correspond to different quantities, in the case of the geochemical
data on nuclear transitions, and the spectroscopic data depends on
hyperfine intervals, only under certain model-dependent
assumptions can they deliver their constraints on $\alpha$
variation. The values related to the strong interaction (a
position of samarium resonance, $g$ factors of nuclei and the
proton mass) should be interpreted in more fundamental terms.
Using certain further assumptions it could be done in terms of the
variation of the dimensionless parameter $m_q/\Lambda_{\rm QCD}$,
where $m_q$ is the quark mass and $\Lambda_{\rm QCD}$ is the
strong interaction QCD scale \cite{Shuryak,Lein}. Strong
constraints on the variation of the electron-to-proton mass ratio and
consequently on parameters of the strong interaction can be
reached from spectroscopy of molecular ro-vibrational levels which
have experienced a substantial progress recently (see., e.g.
\cite{mol1,mol2,mol3}).

The laboratory results on optical measurements in terms of the
effective rate $\partial \alpha/\partial t$ are the least strong
in the list above, but the most reliable. At present these
results show great promise since they are related to optical
clocks, which have progressed rapidly over the past decade.
Recently transitions in a few atoms have been measured accurately,
 and four of them have been studied at
least twice with time separation of few years. Various transitions
were studied in hydrogen \cite{hyd,newH}, calcium
\cite{udem,newCa}, strontium ion Sr$^+$ \cite{srplus}, neutral
strontium \cite{sr1,sr2}, indium ion \cite{in}, ytterbium ion
Yb$^+$ \cite{yb,newYb,newnewYb} and mercury ion Hg$^+$
\cite{udem,hg}. We expect that most of these transitions will soon
provide us with limits on the size of a possible variation the
fine structure constant at the level of a few parts in $10^{-15}$
per a year.

The estimation of a possible $\alpha$ variation from a
model-independent comparison of only optical transitions is based
on an accurate treatment of the relativistic effects. There are
several kinds of searches which were first applied in
astrophysics. They used to deal with different kinds of
transitions, e.g.  a comparison between the hyperfine structure
(HFS) and gross structure transitions, which involve the
fundamental constants in different ways (see, e.g.,
\cite{cjp,amo}). Trying to compare two different HFS transitions
{\em Prestage et al.\/} \cite{prestage} suggested taking into
account relativistic corrections, that have quite different values
for light and heavy ions. They pointed out that the relativistic
contribution is, in fractional units, of the order of
$(Z\alpha)^2$, where $Z$ is the nuclear charge. Such a big
correction takes place even in a neutral alkali atoms and ions
with a low degree of ionization \cite{casimir}, where the electron
may be expected to see a screened nuclear charge much smaller than
$Z$. This happens because the correction chiefly originates not
from a broad area far from nucleus, but from a narrow area close
to it. {\em Dzuba et al.} \cite{dzubaprl} applied this idea to
optical transitions and developed a more accurate quantitative
theory for transitions of hyperfine, fine and gross structure for
most atoms of metrological and astrophysical interest
\cite{dzuba,dzuba1}. They also pointed out \cite{dzuba1} that it
may happen that the non-relativistic term (which is of order of
$Ry$) and the relativistic contribution ($\alpha^2 Ry$) may
accidentally nearly cancel each other, and two states with
different non-relativistic structure (e.g. with a different
orbital number $L$) can have nearly the same energy. In particular
they suggested a measurement of the transition frequency between
two states of dysprosium which both have energy of
19797.97~cm$^{-1}$, the same total momentum $J=10$ but opposite
parity. They belong to the $4f^{10}5d6s$ and $4f^95d^26s$
configurations. The experiment with Dy is now in progress
\cite{budker}.

The frequency of a transition between two atomic states can be
presented in the form
\begin{equation}
  f\simeq c_1{\rm Ry} + c_2(Z\alpha)^2 {\rm Ry},
\end{equation}
where $Ry$ is Rydberg constant in frequency units, and $c_1$ and
$c_2$ are coefficients representing the size of the
non-relativistic and relativistic terms respectively and $Z$ is
the nuclear charge. The sensitivity of the frequency to variation
of $\alpha$ can be described by a value
\begin{equation} \kappa = \frac {\partial \ln \bigl(f/{\rm
Ry}\bigr) }{\partial \ln \alpha},
\end{equation}
which relates change of $\alpha$ to change of frequency
\begin{equation}
\frac {\partial \ln\bigl(f/{\rm Ry}\bigr)}{\partial t}=\kappa
\frac {\partial \ln\alpha}{\partial t}\;.
\end{equation}
One can show that
\begin{equation}\label{kappa}
\kappa \simeq \frac{2 c_2(Z\alpha)^2}{c_1 + c_2(Z\alpha)^2}\;.
\end{equation}

In most of situations the coefficients $c_1$ and $c_2$ are both of
order of unity. In this case for light atoms (low $Z$) the
sensitivity is about or below
\[
\kappa = O((Z\alpha)^2)\;.
\]
For higher $Z$ when $Z\alpha$ is not a small parameter anymore
$\kappa$ is $O(1)$. However, there may be a specific situations
with the denominator in Eq. (\ref{kappa}) close to zero and in
such cases $\kappa$ may be much larger than unity delivering an
enhancement factor. In this paper we consider the possibility of
performing precision experiments with neutral atoms and
singly-charged ions with high $\kappa$ values, that may range from
figures substantially below unity up to $10^8$ (as it is for
dysprosium \cite{budker}). However, advantages of this great
enhancement in the latter atom are of reduced value since one of
these Dy levels is relatively broad, resulting in a measurement of
the splitting between the degenerate levels being limited to a
certain fraction of the linewidth. In \cite{cjp} a slightly
different idea was suggested, namely not to limit a search to only
levels with very big enhancement, but instead to require narrow
levels.

In this paper we present several examples of narrow transitions with
enhanced sensitivity. We consider a possible enhancement of
sensitivity to $\alpha$ variation, and pay special attention to
the feasibility of a high resolution spectroscopic experiment, this
implies a number of additional conditions on the spectrum.

A big value of the enhancement factor $\kappa$ obviously increases
the sensitivity of a transition frequency to a possible $\alpha$
variation. The consequences can be clearly seen from the identity
\begin{equation}
\frac{\Delta f}{f} = \kappa \frac {\Delta \alpha}{\alpha}\;,
\end{equation}
where we suggest that a variation of frequency, $f$, can be
expressed as a variation of $\alpha$, with a variation of the
Rydberg constant neglected.
The latter is possible because current laboratory constraints on
the possible variation of $\alpha$ and the numerical value of the
Rydberg frequency, $Ry$, are at the same level \cite{newYb,amo}
(since they were obtained in atomic systems with $\kappa=O(1)$).

The relativistic effects can in principle strongly affect the
non-relativistic theory. However the relativistic contributions to
a transition frequency related to the gross structure cannot be
enormously big and thus, cannot alone be responsible for a big
enhancement. Actually, such transitions are possible, namely the
transitions between fine structure components, for which the
non-relativistic term is equal to zero. In fact this does not help
much with sensitivity, it can be easily seen from Eq.
(\ref{kappa}) that this gives $\kappa=2$.

The origin of a large enhancement is a strong cancellation between
the non-relativistic and relativistic terms which drastically
reduces the value of the frequency. Both the non-relativistic
terms and the relativistic contributions have, for each atomic
system, certain characteristic values that set margins on possible
$\kappa$ values, these typically cannot exceed the level of few
units of $(Z\alpha)^2$. A really big enhancement factor $\kappa$
may appear if the $\alpha$ dependence of the frequency still has a
characteristic value (in absolute units), but the frequency itself
is small (i.e. the denominator of Eq. (\ref{kappa}) is small). The
widths of the levels also have certain typical values in each
atom, varying for different kinds of transitions and due to
external effects. For the most narrow lines effects due to
collisions or residual external field may be dominant in the real
linewidth.

Summarizing, we note that the transitions with a high sensitivity,
$\kappa$, should possess low frequencies, but we can only take
advantages of their sensitivity if the levels are narrow enough.
If the level has a low frequency but a typical linewidth, the
fractional uncertainty goes up. Only with the narrow lines can we
hope to reach a high relative accuracy. Otherwise, a cancellation
will lead to an enhancement of the sensitivity and simultaneously
to a reduction of a fractional accuracy by approximately the same
factor.

Currently, development of highly-accurate frequency standards
involves transitions with higher and higher frequency and, in
particular, optical transitions in neutral atoms or slightly
charged ions. The use of optical frequencies potentially allows one to
achieve a higher accuracy because of a much larger number of
oscillations in a given time compared with microwave frequency
standards. Choosing optical lines with small natural linewidths in
general also reduces the relative influence of different
systematic effects on the transition frequency and, as a result,
on the accuracy of an optical standard.

At present, a number of frequency standards, based on narrow
optical transitions in  neutral and singly-ionized atoms are
considered as the candidates for a new generation of the frequency
standards with an extremely high level of accuracy
\cite{Hollberg,newnewYb,newH,udem,newCa,srplus,sr1,sr2}. The level
of fractional uncertainty $\Delta f/f$ of the best measurements up
to now has been a few parts in $10^{-15}$, however, estimations of
the possible accuracy of the presently discovered optical
frequency standards give an upper limit much better than the
present results approaching the level of $10^{-18}$. Interest in
the development and application of optical frequency standards for
fundamental physics experiments has been stimulated during the
last several years due to the invention of the optical frequency
comb synthesizer \cite{comb}, which provides a simple and
extremely accurate link between the optical and radio frequency
(RF) domain.

In spite of the recent progress in the development of the optical
frequency standards, the Cs radio-frequency standard, with the
transition frequency of about 0.3 cm$^{-1}$ still remains the most
accurate. The use of slow rubidium atoms can allow further
improvement to the frequency stability of such RF standards, due
to smaller collision frequency shifts, and can potentially reach
the quantum limited level of $\Delta f/f \sim 10^{-16}$
\cite{paris}. Thus, on the level exceeding the accuracy of the RF
standards ($10^{-16}$ and higher), highly accurate optical
frequency standards must be compared directly with each other. On
such a level of accuracy this can be realized only by bringing the
optical standards to one place, which would lead to serious
experimental difficulties, such as  creation of transportable
standards possessing an ensured extremely high level of accuracy.

\section{Transitions with a narrow linewidth and an enhanced
sensitivity to $\alpha$ variation}

A promising approach to overcome some of these difficulties
involves the  creation of a frequency standard of even moderate
accuracy, but based upon a transition with a large relativistic
correction. This would allow the performance of highly-sensitive
experiments with the aim of placing tight constraints upon a
possible $\alpha$ variation, by comparing a possibly
time-dependant frequency with the well developed Cs frequency
standards, linked by GPS to the primary Cs standards at, e.g.,
NIST or PTB. A number of possible candidates for a frequency
standard, with large enhancement factors for a possible detection
of $\alpha$ variation, are listed in Tables~\ref{tab1} and
\ref{tab2}.

Accurate relativistic calculations are needed to reveal how an
energy level will change with time in the presence of $\alpha$
variation. Following \cite{dzuba1}, we represent the energy of a
level by
\begin{equation}\label{eq:omega}
\omega = \omega_{0} + q x
\end{equation}
where $x = (\alpha/\alpha_{0})^{2}-1$, $\omega_{0}$ is the initial
value of $\omega$ (i.e. the one measured at the beginning of the
experiment) and $q$ is a coefficient that determines the frequency
dependence on a variation of $\alpha$. Then the enhancement factor
$\kappa$ (\ref{kappa}) becomes
\begin{equation}\label{kappaq}
\kappa = \frac{2\Delta q}{\Delta \omega_{0}}
\end{equation}
where $\Delta q = q_{2}-q_{1}$ represents the difference in the
$q$ coefficients, and $\Delta \omega_{0}$ the difference in
energy, of the levels between which the transition occurs.

The measurement of the energy shift between two levels will be
easiest to measure when this energy shift is large. It follows
that the best situation would be to have two levels with
relatively large shifts but with opposite sign. These levels would
drift apart relatively rapidly as time passed. A compromise would
be to find two levels with very different $q$ coefficients. A
level with a small $q$ coefficient will not stray much from its
initial value while a level with a large $q$ coefficient will move
quite fast, the first level can then act as a reference point for
the movement of the second level. As a rule of thumb the $q$
coefficients are negative for $s_{1/2}$ and $p_{1/2}$ states and
positive for other states \cite{dzuba}. The easiest way to ensure
that two states have different $q$ coefficients is to ensure that
they have substantially different electron configurations.

In Table~\ref{tab1} we list pairs of long-lived almost degenerate
states of different configurations. Here enhancement is mostly due
to the small energy interval between the states. However, the fact
that the configurations are different also contribute to the
enhancement. Most of the transitions presented in the table
correspond to $s - d$ or $d - f$ single-electron transitions.
Since relativistic energy shifts $q$ strongly depend on $l$ and
$j$ of individual electrons~\cite{dzuba} it is natural to expect
that $\Delta q$ is large for the transitions.

In Table~\ref{tab2} we list some metastable states that are close
to the ground state. Here the enhancement is smaller due to larger
energy intervals. However, measurements would be easier to perform
due to convenience of dealing with transitions from the ground
state.

Enhancement factors $\kappa$, presented in Tables~\ref{tab1} and
\ref{tab2} are calculated in a singe-electron approximation which
doesn't take into account configuration mixing. These calculations
can be considered as rough estimations only. Configuration
interaction can significantly change the values of $\kappa$ in
either way. For example, states of the same parity and total
momentum $J$ separated by small energy interval are likely to be
strongly mixed. Therefore, the assignment of these states to
particular configurations is ambiguous and the relative value of
the relativistic energy shift $\Delta q $ is likely to be small.
An enhancement factor $\kappa $ for such states is difficult to
calculate. Its value is unstable because the transition frequency
$\Delta \omega_{0}$ is also small. We do not include pairs of
states of the same parity and momentum in Table~\ref{tab1}. One
can still find metastable states of the same parity and total
momentum as the ground state in Table~\ref{tab2}. Here mixing of
states can be small due to the large energy separation between the
states.

States of the same parity but different total momentum $J$ can be
affected by configuration mixing in a very similar way. They can
be mixed with states of appropriate values of $J$ from other
configurations. This would also bring values of $q_1$ and $q_2$
for two states closer to each other. On the other hand,
configuration mixing can cause anomalies in fine
structure~\cite{Dzuba05} or in general can have different effect
on different states within the same configuration which would lead
to increased sensitivity of the energy intervals to the variation
of $\alpha$. The detailed study of the enhancement in each listed
transition goes far beyond the scope of the present work. It can
be done in a much more detailed and accurate way during the
planning stage of a specific experiment.

\begin{table*}
\caption{Long lived almost degenerate states with large
$\kappa=\frac{2\Delta q}{\Delta \omega_{0}}$, where $\Delta q =
q_{2}-q_{1}$.} \label{tab1}
\begin{ruledtabular}
\begin{tabular}{lc llcr llcr c }
\multicolumn{1}{c}{Atom} & & \multicolumn{4}{c}{First State}
& \multicolumn{4}{c}{Second State}& $\kappa$ \\
\multicolumn{1}{c}{or ion} & $Z$ &
\multicolumn{2}{l}{Configuration} & $J$ &
\multicolumn{1}{c}{Energy (cm$^{-1}$)}
& \multicolumn{2}{l}{Configuration} & $J$ & \multicolumn{1}{c}{Energy (cm$^{-1}$)} &  \\
\hline Ce~I  & 58 & $4f5d^{2}6s$ & $^5$H &  3  &  2369.068 &
             $4f5d6s^{2}$ & $^1$D &  2  &  2378.827 & -2000 \\

      &    & $4f5d^{2}6s$ &       &  4  &  4173.494 &
             $4f5d6s^{2}$ & $^3$G &  5  &  4199.367 & -770 \\

      &    & $4f^26s^{2}$ & $^3$H &  4  &  4762.718 &
             $4f5d6s^2$   & $^3$D &  2  &  4766.323 & -13000 \\

Ce~II & 58 & $4f5d6s$     & $^4$F & 9/2 &  5675.763 &
             $4f5d^{2}$   &       & 7/2 &  5716.216 &  500   \\

      &    & $4f5d^{2}$   & $^4$S & 3/2 &  8169.698 &
             $4f5d6s$     & $^4$D & 5/2 &  8175.863 & -3300    \\

Nd~I  & 60 & $4f^35d6s^2$ & $^5$K &  6  &  8411.900 &
             $4f^45d6s$   & $^7$L &  5  &  8475.355 &  950   \\

      &    & $4f^35d^26s$ & $^7$L &  5  & 11108.813 &
             $4f^45d6s$   & $^7$K &  6  & 11109.167 & $10^5$ \\

      &    & $4f^45d6s$   & $^7$I &  7  & 13798.860 &
             $4f^35d^26s$ & $^7$K &  7  & 13799.780 & $-4\cdot 10^4$ \\

Nd~II & 60 & $4f^{4}5d$   & $^6$L & 11/2 &  4437.558 &
             $4f^{4}6s$   & $^4$I & 13/2 &  4512.481 &  -270    \\

      &    & $4f^{4}5d$   & $^6$G & 11/2 & 12021.35 &
             $4f^{4}6s$   & $^6$F & 9/2 & 12087.17 &  -300    \\

Sm~I  & 62 & $4f^66s^2$   & $^5$D &  1  & 15914.55 &
             $4f^{6}5d6s$ & $^7$G &  2  & 15955.24 &  500    \\

Eu~I  & 63 & $4f^{7}6s6p$ & $^{10}$P & 11/2 & 15581.58 &
             $4f^{7}5d6s$ & $^8$D    & 9/2 & 15680.28 &  100  \\

Gd~II & 64 & $4f^{7}5d6s$ & $^8$D    & 11/2 &  4841.106 &
             $4f^{7}5d^2$ & $^{10}$F & 9/2 &  4852.304 &  1800  \\

      &    & $4f^{7}5d^2$ & $^{10}$P & 7/2 & 10599.743 &
             $4f^{7}5d6s$ & $^6$D    & 5/2 & 10633.083 &  -600   \\

Tb~I  & 65 & $4f^96s^2$   & $^6$H    & 13/2 &  2771.675 &
             $4f^85d6s^2$ & $^8$G    & 9/2 &  2840.170 &  -600   \\

Tb~II & 65 & $4f^85d6s$   &          &  6  &  5147.23~ &
             $4f^96s$     &          &  6  &  5171.76~ &  1600   \\

\end{tabular}
\end{ruledtabular}
\end{table*}

\begin{table*}
\caption{Metastable states sensitive to variation of $\alpha$.}
\label{tab2}
\begin{ruledtabular}
\begin{tabular}{lc llc llcr c }
\multicolumn{1}{c}{Atom} & & \multicolumn{3}{c}{Ground State}
& \multicolumn{4}{c}{Metastable State}& $\kappa$ \\
\multicolumn{1}{c}{or ion} & $Z$ &
\multicolumn{2}{l}{Configuration} & $J$
& \multicolumn{2}{l}{Configuration} & $J$ & \multicolumn{1}{c}{Energy (cm$^{-1}$)} &   \\
\hline
La~I  & 57 & $5d6s^2$   & ~$^2$D & 3/2 &  $5d^26s  $ & $^4$F    & 5/2 &  3010.002 &  6.6 \\
      &    &            &       &     &  $5d^3    $ & $^4$F    & 3/2 & 12430.609 &  3.2 \\
La~II & 57 & $5d^2$     & ~$^3$F &  2  &  $5d6s$     & $^3$D    &  1  &  1895.15~ & -10  \\
      &    &            &       &     &  $6s^2    $ & $^1$S    &  0  &  7394.57~ &  -5.4 \\
Ce~II & 58 & $4f5d^2  $ & ~$^4$H & 7/2 &  $4f5d6s  $ &          & 9/2 &  2382.246 &  -8 \\
Pr~I  & 59 & $4f^36s^2$ & ~$^4$I & 9/2 &  $4f^35d6s$ & $^6$L    & 11/2 &  8080.49~ & 2.5 \\
Pr~II & 59 & $4f^36s  $ &       &  4  &  $4f^35d  $ & $^5$L    &  6  &  3893.46~ &  5 \\
Nd~I  & 60 & $4f^46s^2$ & ~$^5$I &  4  &  $4f^45d6s$ & $^7$L    &  5  &  8475.355 & 2.6 \\
Nd~II & 60 & $4f^46s  $ & ~$^6$I & 7/2 &  $4f^45d  $ & $^6$L    & 11/2 &  4437.558 & 4.5 \\
Sm~I  & 62 & $4f^66s^2$ & ~$^7$F &  0  &  $4f^65d6s$ & $^9$H    &  1  & 10801.10~ &  2  \\
Sm~II & 62 & $4f^66s  $ & ~$^8$F & 1/2 &  $4f^65d  $ & $^8$H    & 3/2 &  7135.06~ &  3   \\
Eu~I  & 63 & $4f^76s^2$ & ~$^8$S & 7/2 &  $4f^75d6s$ & $^{10}$D & 5/2 & 12923.72~ &  1.9   \\
Eu~II & 63 & $4f^76s  $ & ~$^9$S &  4  &  $4f^75d  $ & $^{9 }$D &  2  &  9923.00~ &  2   \\
Gd~I  & 64 & $4f^75d6s^2$ & ~$^9$D  &  2  &  $4f^75d^26s$ & $^{11}$F & 2 &  6378.146 & 3   \\
Gd~II & 64 & $4f^75d6s$ & $^{10}$D & 5/2 &  $4f^76s^2$ & $^{8}$S & 7/2  &  3444.235 & -6   \\

Tb~I  & 65 & $4f^96s^2$ & ~$^{6 }$H & 15/2 &  $4f^85d6s^2$ & $^8$G & 13/2  &   285.500 &  -140 \\
Pt~I  & 78 & $5d^96s$ & ~$^3$D &  3  &  $5d^86s^2$ & $^4$F &  4   &   823.7~~ &  -24 \\
Pt~II & 78 & $5d^9  $ & ~$^2$D & 5/2 &  $5d^86s  $ & $^4$F & 9/2  &  4786.6~~ &   -6 \\
Ac III& 89 & $7s    $ & ~$^2$S & 1/2 &  $6d      $ & $^2$D & 3/2  &   801.0~~ &  25 \\
      &    &          &       &     &  $6d      $ & $^2$D & 5/2  &  4203.9~~ &   5 \\
\end{tabular}
\end{ruledtabular}
\end{table*}

\section{Consideration of particular candidates}

For several atoms from Table~\ref{tab2} we made some rough
estimates on the practical realization as candidates for possible
frequency standards. Basic criteria, in spite of general lack of
available information, included lifetime of the clock transition
as well as the possibility of detecting the excitations and
cooling the atoms in order to reduce a number of systematic
frequency shifts, such as second-order Doppler effect, collision
shifts, etc.

\subsection{Neutral platinum and ion Pt$^+$ (Pt II)}

The spectrum of neutral and singly ionized platinum is attractive
for a search for a possible $\alpha$ variation because of the
relatively large relativistic corrections (see Table~\ref{tab2}).
According to \cite{Pt_lifetime}, the lifetimes of the $^3$F$_{4}$
and the $^4$F$_{9/2}$ levels of Pt and Pt$^+$ are extremely high,
corresponding \textit{gA} values (\textit{g} is degeneracy of the
level, \textit{A} is the Einstein spontaneous transition rate) are
of the order of 10$^{-9}$ s$^{-1}$. The use of the stable
$^{195}$Pt isotope with the nuclear spin I = 1/2 would increase
the transition dipole moment due to nuclear spin-orbit
interaction, however leading to additional re-pumping from the HFS
sublevels. Because of the rich energy structure,
 direct laser cooling of Pt
and Pt$^+$ is difficult to realize. The use of the appropriately
strong E1 transitions from the 5d$^7$6s~$^4$F$_2$ in Pt as well as
a similar transition from the 5d$^8$6p~$^4$D$_{7/2}$ in Pt$^+$
would require a large number of re-pumping lasers, which increases
the complexity of the setup. In the case of Pt$^+$ one can
consider trapping of a single platinum ion in a quadrupole
radio-frequency trap and the use of the sympathetic cooling
approach \cite{Sympath_cooling} to reduce a temperature of the
ion. This can also provide an efficient excitation detection on
the Pt$^+$ clock transition via a vibration motion phonon exchange
with a coolant ion (i.e., via so-called ``quantum-logic''
spectroscopy) \cite{Wineland_sympath_standard}.

\subsection{Actinium ion Ac$^{++}$ (Ac III)}

The francium-like doubly ionized actinium ion possesses a
relatively simple spectrum. There is no information about lifetime
of the $6d_{3/2}$ and the $6d_{5/2}$ levels, however, due to the
small transition frequency and the same parity of the ground and
the upper levels we estimate the linewidth of the transitions to
be sufficiently small.

Recent results from trapping and high resolution spectroscopy of
the neutral francium atoms \cite{Francium_trapping} in the
magneto-optic traps (MOT) raised interest and opened new
possibilities in the precision study of the radioactive elements,
especially in the tests of the standard model via parity violation
experiments. In application to an optical frequency standard one
can consider trapping a single Ac$^{++}$ ion in a quadrupole
radio-frequency trap. Efficient trapping of the multiply charged
ions has been realized (see, for example, \cite{ref4}), also
single hydrogen-like ions were studied in the Penning trap
\cite{ref5}. However, as in the case of platinum, direct laser
cooling of the actinium ion due to its energy level structure
seems to be a problem and most probably sympathetic laser cooling
and ``quantum-logic'' detection is required.

\subsection{Neutral terbium (Tb I)}

In spite of the absence of information about the lifetimes of the
metastable states $^8$G$_{13/2}$ and $^8$G$_{1/2}$ at 285.5
cm$^{-1}$  in the neutral terbium, we estimate the lifetime to be
large especially for the 8G$_{1/2}$ level, from which transition
to the ground state is strongly forbidden ($J=15/2$ to $J=1/2$
transition). The main natural isotope of terbium ($^{159}$Tb)
possesses nuclear spin $I=3/2$, thus allowing the use of the
advantages of the $ m=0 \to  m^\prime=0$ clock transition.  The
transition wavelength $\lambda=35~\mu$m is quite large which
strongly minimizes the influence of the first- and the
second-order Doppler effect, especially with a reduction of the
effective temperature of the atoms. However, laser cooling of
terbium is not possible due to its very rich energy structure.
According to the Boltzman distribution the metastable
$^8$G$_{13/2}$ state at 285.8 cm$^{-1}$ has a significant thermal
population at room temperature (occupation number is about 0.25)
and some kind of re-pumping should be applied in order to perform
the high-resolution spectroscopy of this transition. A cryogenic
cooling of the apparatus should be used to reduce the influence of
the black body radiation (BBR) on the clock transition by means of
an induced ac-Stark and ac Zeeman shifts \cite{Blackbody_shifts}.
High-resolution spectroscopy on the terbium clock transition
entails the problem of  creation of a coherent radiation source at
the wavelength of 35~$\mu$m with a high level of frequency
stability and spectral purity. Apart from the use of a different
kinds of submillimeter lasers or the quantum cascade lasers
\cite{Quant_casc_las}, another promising approach is the
phase-matched difference-frequency mixing of the frequency stable
optical radiations in certain kinds of nonlinear crystals with a
wide transmitting range (GaP, DAST, \cite{DAST}). Conservation of
the relative frequency stability in the THz radiation by the
optical down-conversion would open a possibility for
high-resolution spectroscopy in the submillimeter range with an
accuracy comparable to that of in the optical measurements. This
would allow to realize the advantages of the low-frequency
transitions with the high $q$-values in the experiments for the
search of a possible time-variation of the fine structure constant
$\alpha$.

\section{Summary}

By itself a large value of the enhancement factor, $\kappa$, is
not enough to develop a highly sensitive search for a variation of
fundamental constants. We list below the necessary conditions.

The general requirement are:
\begin{itemize}
\item Two levels ($A$ and $B$) with different non-relativistic
quantum numbers should be close to each other. The best situation
is related to the case when at least one of valence electrons  is
in the different state,
 e.g., the $s^2$ and $sd$ configurations. However, a different
configuration of electrons in the same electronic states as e.g.
$d^2\;$S and $d^2\;$D is also possible, but the relativistic
corrections in the latter case are smaller. Close levels are
levels where the relativistic separation (fine structure) is
substantially bigger than the difference between the two different
levels $A$ and $B$.

\item The levels must be narrow enough and systematic frequency
shifts on the transition frequency should be small, enabling
accurate determination of the transition frequency. Basically, the
ratio of the relative measurement uncertainty, $\delta f/f$, to
the enhancement factor, $\kappa$, is a characteristic value for
comparison with other searches \footnote{One has to remember that
a standard constraints on $\alpha$ is derived by comparison of two
or more optical transitions with different values of $\delta f/f$
and $\kappa$, which are nevertheless of the same order of
magnitude.}.

\item It should be possible to induce a transition between $A$ and
$B$ and and to have an efficient tool to detect it. Cooling of the
atoms is essential in order to increase the accuracy in frequency
measurements. As we see from our consideration above it is not
easy to satisfy such obvious requirement.

\end{itemize}
The number of successful detection and cooling schemes for
precision spectroscopy is quite limited and this leads to strong
limitations on candidates. However, recent progress in the
"quantum-logic" spectroscopy \cite{Wineland_sympath_standard}
opens  new possibilities in cooling and high-resolution
spectroscopy of a large number of ions.

Advantages of the enhancement are twofold. Firstly, we can make a
measurement with a reduced accuracy and still  reach a competitive
result. This allows one to get rid of certain systematic effects
present in the most precision measurements. Secondly, if a high
precision measurement is possible (as we hope in the case of some
narrow transitions) the enhancement may offer the strongest test
possible in a laboratory study.

In summary, we presented a number of narrow transitions with a
large enhancement factor and discussed various problems involved
in realization of precision frequency measurements for these
transitions. A successful experiment with one of these or,
perhaps, some other similar transitions may set new strong
constrains on a possible variation of the fine structure constant
$\alpha$.

\section*{Acknowledgments}

An initial part of this work was done during a short visit of VF
to MPQ, Garching and he would like to thank MPQ for hospitality.
The work of SK was supported in part by the RFBR under grants
03-02-04029 and 03-02-16843, and by DFG under grant GZ 436 RUS
113/769/0-1. The work of VF was supported in part by the
Australian Research Council and Department of Energy, Office of
Nuclear Physics, Contract No. W-31-109-ENG-38. The authors
gratefully acknowledge stimulating discussions with D. Budker, S.
Schiller and J. L. Hall.

\end{document}